\documentclass[english]{article}
\usepackage[T1]{fontenc}
\usepackage[latin9]{inputenc}
\usepackage{geometry}
\usepackage{graphicx}



\usepackage{babel}

\begin{document}

\title{Parameter degeneracies in FNAL-Homestake LBNE setup}

\author{Kalpana Bora\\
\emph{Physics Department, Gauhati University, Assam, India.}\\
e-mail: kalpana@gauhati.ac.in}

\date{{}}

\maketitle
\begin{abstract}
LBNE (Longbaseline Neutrino Oscillation Experiments) provide a powerful
experimental setup to study sensitivities and exlcusion limits in
neutrino oscillation parameter space. A longbaseline experiment is
being planned, at USA, from FNAL (Fermilab National Accelerator Laboratory)
to an underground laboratory at Homestake in South Dakota, at an angle
of 5.84 degrees from FNAL (at a distance of 1289 km). The prospect
of a new beamline towards this location from FNAL, and a 300 Kiloton
water Cerenkov detector at the site is in planning stage, for the
studies of the neutrino physics program. The long baseline provides
sufficient matter effects for neutrino travel, and a large detrecor
will help towards better statistics. In this work, we present, upto
what extent, the parameter degeneracies, present in oscillation parameter
space, can be resolved, using this FNAL-LBNE setup.
\end{abstract}

\section*{I. Introduction}

There is now sufficient evidence that neutrinos have mass and hence
they oscillate. The experiments with solar {[}1], atmospheric {[}2],
reactor {[}3], and long baseline accelerator neutrinos {[}4], have
provided compelling evidence for the existence of neutrino oscillations.
Recently, MiniBooNE {[}5] and T2K {[}6] have also provided some data.
MiniBooNE {[}5] suggest the evidence of oscillations in the $\bar{\nu_{\mu}\rightarrow}\bar{\nu_{e}}$
sector, at the value $\frac{L}{E}\sim$ 1 km/GeV, where L is the baseline
of the experiment and E is the neutrino energy. This is almost similar
to the result provided by LSND {[}7] about a decade ago. On the other
hand, results from T2K suggests that $\sin^{2}2\theta_{13}>0.03$
{[}6]. This result from T2K boosts the expectation of discovery of
CP violation by some future planned experiments {[}8]. A recent global
analysis of available neutrino data can be found in {[}9]. 

Inspite of the results on values of $\theta_{12},\theta_{23,}\Delta m_{21}^{2},\Delta m_{23}^{2}$
and $\theta_{13}$, there are still some unknowns in neutrino sector
-- mass hierarchy, CP violation phase $\delta_{CP}$, precise value
of $\theta_{13}$ , absolute mass of neutrinos, whether neutrinos
are majorana neutrinos, whether neutrinos constitute a part of Dark
Matter and/or Dark Energy, etc. In this quest to find the unknowns,
long baseline neutrino experiments (LBNE) have a very impotant role
to play. Some of the ongoing and planned LBNEs are - MINOS {[}10],
T2K {[}6], NO$\nu$A {[}11], FNAL-LBNE {[}12] , etc. The LBNEs have
an advantage that due to long baselines, the neutrinos travel long
distances through matter, and hence matter effects become important.
Due to this, they become capable of differentiating between normal
mass hierarchy (NMH, $m_{3}^{2}-m_{2}^{2}>0$) and inverted mass hierarchy
(IMH, $m_{3}^{2}-m_{2}^{2}<0$). This is because matter effects have
opposite signs for these two hierarchies, in the formula for neutrino
oscillation probababilities. Also, if the detector mass is very high
(hundreds of kilotons), then the statistics become better, and preicision
physics become possible. They also become senesitive to measurement
of the unknowm mixing angle $\theta_{13}$, and CPV phase $\delta_{CP}$. 

Along with these accelerator based neutrino physics issues, a very
large detector could also be sensitive to some other studies{[}12].
These are -- improved search for nucleon decay (see {[}39] for latest
limits on proton life time), observation of natural sources of neutrino
(such as the Sun, Earth's atmosphere, Supernova explosion etc.). Also,
there may be galactic sources of neutrinos, galactic neutrinos have
a natural source in inelastic nuclear collisions through leptonic
decays of charged secondary pions. Such neutrino sources, currently
not detectable, could be seen by a large megaton neutrino detector
that runs for several decades.

The LBNEs, although being very useful to study above discussed physics,
suffer from a serious drawback -- the presence of parameter degeneracies,
see ref {[}13-19]. Due to the inherent structure of three flavor neutrino
oscillation probabilities, for a given experiment, in general several
disconnected regions in multi-dimensional space of oscillaion parameters
will be present. So, it becomes difficult to pin-point, which one
is the exact (true) solution. There degeneracies can be classified
as :

\begin{enumerate}
\item \emph{The intrinsic or ($\delta_{CP},\theta_{13})$-degeneracy }{[}20,21]--
As can be observed from the formula for oscillation probability for
the appearance channel $\nu_{\mu}\rightarrow\nu_{e}$, for neutrinos
and anti-neutrinos, for three flavor case, two disconnected regions
appear in the \emph{($\delta_{CP},\theta_{13})$ }plane. But, for
the experiments operating at first oscillation maximum, the second
solution can be disfavored {[}13,15].
\item \emph{The Hierarchy or sign ($\Delta m_{31}^{2})$-degeneracy} {[}22]--
The two degenerate solutons corresponding to two signs of $\Delta m_{31}^{2}$
appear at different values of $\delta_{CP}$ and $\theta_{13}$.
\item The octant or $\theta_{23}$-degeneracy {[}23]-- LBNEs are senesitive
mainly to $sin^{2}2\theta_{23}$, it is difficult to distinguish the
two octants $\theta_{23}<\frac{\pi}{4}$ and $\theta_{23}>\frac{\pi}{4}$.
The solutions corresponding to $\theta_{23}$ and $\frac{\pi}{2}-\theta_{23}$
appear at different values of $\delta_{CP}$ and $\theta_{13}$.
\end{enumerate}
This leads to an eight-fold ($2\times2\times2$) degeneracy, and hence
ambiguity, in the determination of the oscillation parameres $\delta_{CP}$
and $\theta_{13}$. This in turn poses a serious problem, and somehow,
we have to tackle this, to find out these parameters exaclty. Several
methods to resolve these degeneracies have been proposed :

\begin{enumerate}
\item Combination of experiments at various baselines and/or $\frac{L}{E}$
values {[}13,22,24-27].
\item Use of spectral information {[}14, 28].
\item Combination of $\nu_{e}\rightarrow\nu_{\mu}$ and $\nu_{e}\rightarrow\nu_{\tau}$
oscillation channels {[}29].
\item Combination of LBNE and reactor experiments {[}30-34].
\item Combination of LBNE and atmospheric experiments {[}35, 36].
\end{enumerate}
In this work, we present results on the presence of octant degeneracy,
in FNAL-DUSEL setup (1300 km baseline), for a 120 GeV proton beam
from NUMI. However, we have not attempted on, how to resolve them.
But, we expect that, they could be resolved by combining this accelerator
LBNE data, with the atmospheric neutrino oscillation experiment ,
if the same 300 kton water cerenkov detector could in future, be used
for the atmospheric experiment as well. This is because, the atmospheric
neutrino data is sensitive to mass hierarchy (MH), $\theta_{13}$
and octant of $\theta_{23}$, while LBNEs are sensitive to $\theta_{13}$,
$\delta_{CP}$ and $|\Delta m_{31}^{2}|$. Hence, the two can be combined
to break the degeneracies.

The paper has been organized as follows. Section II contains technical
details of the planned FNAL-DUSEL LBNE and the detector (as used in
{[}12]), and the channels being used etc. Section III contains our
results on parameter degeneracies in this experiment. This section
is the highlight of this work. We have used the software GLoBES {[}37]
in our work. Conclusions have been presented in Section IV. A more
detailed analysis of this work will be presented elsewhere {[}38].

\section*{II. Details of the Experiment}

In this section, we will present the technical details of the experiment
and the detector, being used in this work, in general. They have been
used from ref. {[}12], and we are mentioning them here, for the sake
of brevity and completeness of this work. The FNAL-DUSEL beamline
that we have considered here, is pointing from NuMI (Neutrino Main
Injector) to Homestake mine in South Dakota, at an angle of $0.5^{o}$.
The baseline from FNAL to Homestake is 1298 (1300) km. The 120 GeV
proton beam is from Fermi-Lab accelerator only, which interacts with
the target to produce muons. These muons then decay and produce muon-neutrinos. 

The detector is a 300 kiloton water Cerenkov detector. Running time
is 3 years for neutrinos and 3 years for anti-neutrinos. Base line
is 1300 Kilometers (FNAL-DUSEL), the proton beam is 120 Gev, 0.5 degrees
off-axis%
\footnote{Prof. Mary Bishai has been kind enough to send me the 120 GeV proton
beam flux files (back in 2010), (that have been used in ref {[}12]
in figs 9-13), via e-mail.%
}. Signals used are $\nu_{\mu}$ and $\bar{\nu}_{\mu}$ appearance
channels, and the background used are NC and electron-beam events.
The cross-sections used for both are as available with GLoBES software,
i.e. those used with NOvA experiment. Energy window used for the analysis
is 0.5-12.0 GeV. The pre- and post-smearing effieciencies, that have
been used are taken from {[}12], which is also the spectral information.
Here, the bin-wise efficiencies have been used, both for the signal
and the background. The earth matter profile used is type 1 of GLoBES
(constant density). Energy resolution used is 10 \% for the electrons,
and 5 \% for the muons, 1\% systematic error on the signal, and 10
\% systematic error on the background has been used. The true values
of the parameters used are as follows:

\[
\sin^{2}(2\theta_{12})=0.86\]

\[
\theta_{23}=\pi/4\]

\[
\Delta m_{21}^{2}=0.86\times10^{-5}\]

\[
\Delta m_{31}^{2}=2.7\times10^{-3}\]

where $\Delta m^{2}$s are in $eV^{2}$ and, 10 \% error on solar
mass difference, 5 \% error on atomspheric mass difference, 5 \% error
on atomspheric angle, 10 \% error on solar angle, 5 \% error on earth
matter density has been used.

\section*{III. Results and Analysis on Neutrino Mixing Parameter Degeneracies}

Using the information given in Section II., we have generated the
contours showing the parameter degeneracy in neutrino oscillation
parameter space. Our results have been presented in figures {[}1-3].
In these figures, true $\delta_{cp}=153^{o}$ , and systematics errors
used are, 10\% error in solar parameters and 5\% error in atomspheric
parameters. Following observations are worth mentioning:

\begin{enumerate}
\item All these figures are drawn at $3\sigma$ (99.73 \%) CL for two parameters
(i.e. at $\chi^{2}=11.83$ ).
\item The true parameters used are :\[
\sin^{2}(2\theta_{12})=0.86,\sin^{2}\theta_{23}=0.6,\delta_{cp}=153^{o}\]

\item Inclusion of systematics makes allowed regions bigger.
\item At $\sin^{2}2\theta_{13}=0.007,$ 4-fold degeneracy can be seen, while
at $\sin^{2}2\theta_{13}=0.03$ and $0.01$, only 2-fold (Octant)
degeneracy is seen to be present.
\item The allowed regions become bigger as $\sin^{2}2\theta_{13}$ becomes
small. It means, it becomes difficult to pinpoint the actual value
of the parameter, or we can say, as $\sin^{2}2\theta_{13}$ decreases,
more difficult it becomes to resolve the degeneracy. 
\end{enumerate}
These results can be further anlysed as follows:

\begin{itemize}
\item Fig 1 - Here, only two degenerate solutions for TH-TO (green), TH-WO
(blue) can be seen, but separation among them is not much. It means
that mass hierarchy has been resolved at $\sin^{2}2\theta_{13}=0.03$,
but Octant degeneracy is present. The green (TH-TO) and blue (TH-WO)
occur at $\sim$ same value of $\delta_{cp}$, i.e. octant degeneracy
affects measurement of $\sin^{2}2\theta_{13}$, while it does not
affect measurement of $\delta_{cp}$. In other words, we can pinpoint
the value of $\delta_{cp}$ even in the presence of octant degeneracy,
while the value of $\theta_{13}$ cannot be pinpointed. \emph{So,
we definitely need another mechanism to break the Octant degeneracy.}\textbf{\emph{
}}.\textbf{ }%
\begin{figure}
\begin{centering}
\includegraphics[width=0.8\textwidth]{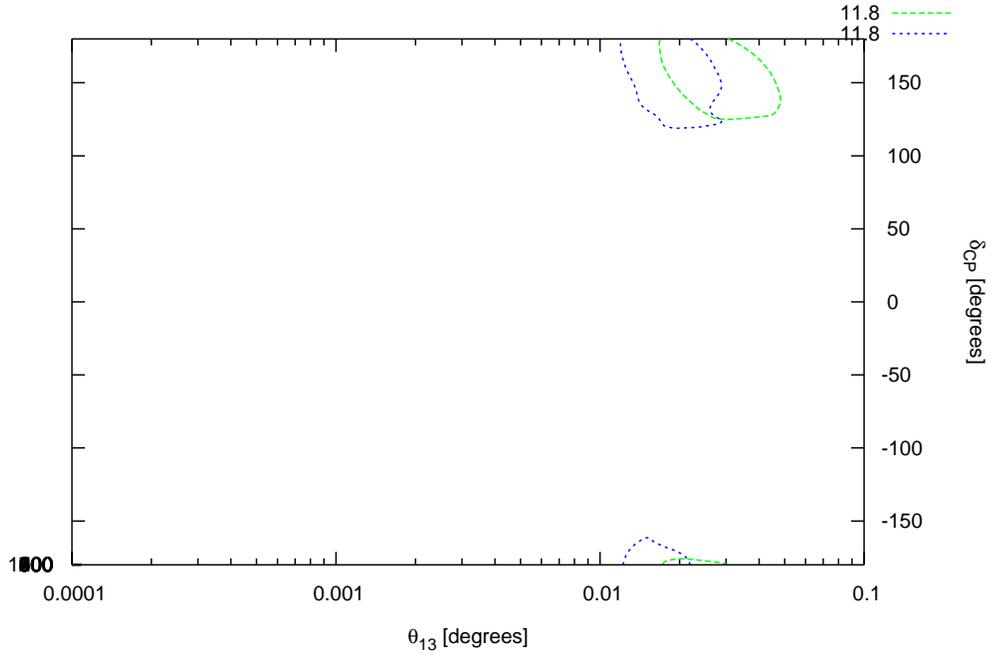}
\par\end{centering}

\textbf{\caption{3$\sigma$ CL contours, showing allowed regions of true and degenerate
solutions for $\sin^{2}2\theta_{13}^{{\rm true}}$=0.03, and $\delta_{CP}^{{\rm true}}=0.85\pi$,
for true NH. Blue (dark) curve is for TH-WO, and green (light) for
TH-TO.}
}
\end{figure}

\item Fig 2 - Here again, the situation is similar to $\sin^{2}(2\theta_{13})=0.03$
(fig 1 above), only allowed regions are bigger, i.e. it becomes little
\emph{more difficult to pinpoint the exact value of the true parameters}.
Also, the TH-TO (green) and TH-WO (blue) curves are seen to be entangled,
it means that it will be even more difficult to break the Octant degeneracy.
Here also, we definitely need another mechanism to break the Octant
degeneracy. %
\begin{figure}
\begin{centering}
\includegraphics[width=0.8\textwidth]{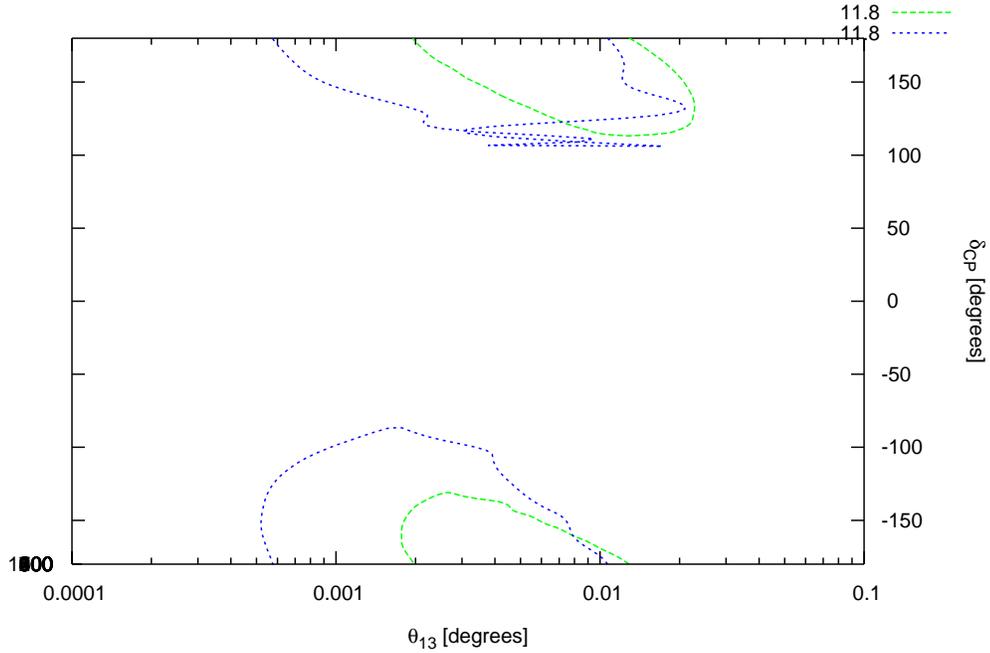}
\par\end{centering}

\caption{3$\sigma$ CL contours, showing allowed regions of true and degenerate
solutions for $\sin^{2}2\theta_{13}^{{\rm true}}$=0.01, and $\delta_{CP}^{{\rm true}}=0.85\pi$,
for true NH. Blue (dark) curve is for TH-WO, and green (light) for
TH-TO.}

\end{figure}

\item Fig 3 - The situation here is similar to fig 2 above, only the allowed
region is even bigger, i.e. it \emph{is even more difficult to pinpoint
the exact value of the parameters}. Here, all the four degenerate
solutions, i.e. TH-TO (green), TH-TO(blue), WH-TO (pink), WH-WO (cyan),
are present. So, it will be most difficult here to pinpoint the true
solution, as four-fold degeneracy is present.%
\begin{figure}
\begin{centering}
\includegraphics[width=0.8\textwidth]{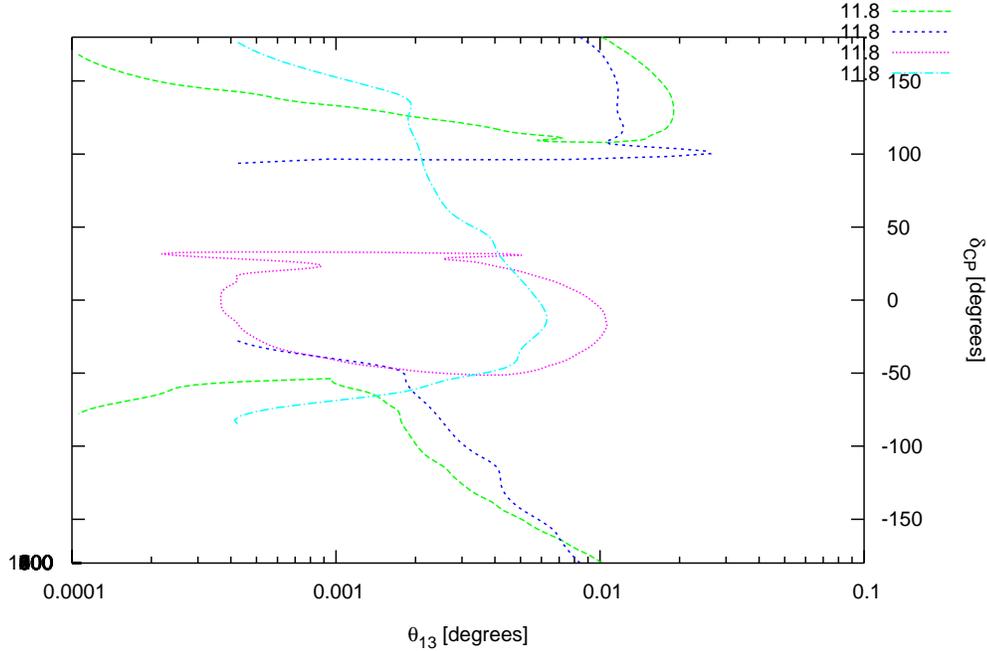}
\par\end{centering}

\caption{3$\sigma$ CL contours, showing allowed regions of true and degenerate
solutions for $\sin^{2}2\theta_{13}^{{\rm true}}$=0.007, and $\delta_{CP}^{{\rm true}}=0.85\pi$,
for true NH. Blue (dark, small dots) curve is for TH-WO, green (light,
small dots) for TH-TO, pink (continuous) for WH-TO, cyan (big dots)
for WH-WO. }

\end{figure}

\end{itemize}
So, we need information from another experiment, to break the Octant
degeneracy present, even at $\sin^{2}2\theta_{13}=0.03$ and $0.01$.

\section*{IV Conclusions}

To conclude, in this work, we presented results on parameter degeneracies,
of neutrino oscillation parameters, in a proposed FNAL-DUSEL (Homestake)
Long Baseline Neutrino Experiment, for a 1300 km baseline, for a 300
kton water cerenkov detectror. We find that, in this experiment, the
mass hierarchy has been resolved at $\sin^{2}2\theta_{13}=0.01$ and
$\sin^{2}2\theta_{13}=0.03$, so CL contours appear only for true
hierarchy. Two contours appear, for TO and WO, but no WH contour is
present. But at low values of $\sin^{2}2\theta_{13}=0.007$, four-fold
degeneracy is present. In all the three figures, WO contours are present,
so a mechanism is needed to reslove this octant degeneracy. As discussed
in text, combination with atmospheric experiment with same detector,
if such measurements could be performed in future, could help resove
octant degeneracy.

\section*{Acknowledgements}

The author thanks Raj Gandhi, Pomita Ghoshal, S. Uma Sankar and Sushant
Raut for many fruitful discussions, during various stages of this
work. She also thanks Patrick Huber and Walter Winter for dicussion
on GLoBES over e-mail (during 2009 and 2010). Mary Bishai deserves
thanks for providing the flux files (in 2010) for the DUSEL FNAL-LBNE
(1300 km) setup, for the 120 GeV proton beam. She also would like
to thank UGC-SAP program, and HRI (Harish Chandra Research Institute),
Allahabad, for financial support, to visit HRI. Thanks are also due
to TIFR (Tata Institute for Fundamental Research), Mumbai, for finacial
assistance to visit TIFR. Parts of this work have been carried out
at HRI, Allahabad, and TIFR, Mumbai.

\end{document}